\documentclass{svjour3}                     
\smartqed  
\usepackage{graphicx}
\usepackage{graphicx,epsfig}
\usepackage[english]{babel}
\usepackage{subfigure}
\usepackage{graphicx}
\usepackage{latexsym}
\usepackage[colorlinks,ps2pdf]{hyperref}
\usepackage{amsmath}
\usepackage{amssymb}

\begin{document}

\title{The NP right-chiral $CC$ coupling constant estimation in neutrino oscillation experiments
}

\titlerunning{Right-Chiral $CC$ Coupling Constant Estimation in Neutrino Experiments}        

\author{Jacek Syska}


\institute{J. Syska \at
              Institute of Physics, University of Silesia, 75 Pu{\l}ku Piechoty 1, Pl 41-500 Chorz{\'o}w, Poland \\
              \email{jacek.syska@us.edu.pl}
}

\maketitle
\begin{abstract}
The error probability of the discrimination of the Standard Model (SM) with massive neutrinos and its new physics (NP) model extension in
experiments of the muon neutrino oscillation, following
the pion decay $\pi^{+} \rightarrow \mu^{+} + \nu_{\mu}$, is calculated.
The stability of the estimation of the NP
charged current coupling constant $\varepsilon_{R}$
is analysed and the robustness of this estimation is checked.
It is shown that the upper bound on the error probability of erroneous identification of the Standard Model with its NP model extension has reached the significantly small value of approximately $2.3 \times 10^{-6}$.
\end{abstract}

\vspace{-4mm}
\section{The muon neutrino density matrix}
\label{effective Hamiltonian}
\vspace{-1mm}

The well known modelling of the chiral right-handed
currents is connected with
left-right symmetric extensions of the Standard Model (SM) \cite{Siringo,Mohapatra-Pati,Zuber}.
There are also effective-Lagranganian SM extensions which can be used to inspect the existence of the chiral right-handed
interactions \cite{Bergmann-Grossman-Nardi,A-B-Sz-Wudka-Z,AZS}.
This paper follows this path.
Let the muon neutrino $\nu_{\mu}$ be produced in the decay $\pi^{+} \rightarrow \mu^{+} + \nu_{\mu}$ of pion to muon and the muon Dirac neutrino  \cite{Giunti-Kim}.
The neutrino $\nu_{\mu}$ produced in this process
is the relativistic one.
The muon flavour neutrino state $|\nu_{\mu}\rangle$ is
a superposition of the stationary states  $|\nu_{i}\rangle_{\lambda} \equiv |{\rm p}, \lambda, i\rangle \,$  \cite{Giunti-Kim} of de\-finite masses $m_{i}$, $i=1,2,3$, helicities $\lambda=-1$ or $+1$ and four-momentum~${\rm p}$ \cite{OSZ}.
By including new physics (NP) interactions \cite{Kuno-Okada}, e.g., the chiral right-handed interactions \cite{AZS},
this superposition composes the mixed state \cite{OSZ,ZZS}.
%
The other reason of the departure from the pure state
can be connected, e.g., with
the existence of
scalar interactions  \cite{OSZ}.
%
From the $\pi^{+}  \rightarrow \mu^{+}  + \nu_{\mu}$ decay experiments
we know that the fraction of the right-handed
$N_{\nu_{+1}}$ to the left-handed  $N_{\nu_{-1}}$ neutrinos fulfils the constraint $N_{\nu_{+1}}/N_{\nu_{-1}} < 0.002$ \cite{PDG_epsR_2,PDG_epsR_3}.
Let us assume that the pion decays effectively both
in the left ($L$) and right ($R$) chiral charged current ($CC$) interactions \cite{ZZS} via the exchange of the
SM $W$-boson only.
Then, at the $W$-boson energy scale, the $R$ and $L$ chiral pion decay constants
\cite{Berman-Kinoshita-1,Berman-Kinoshita-2,Berman-Kinoshita-3}
are equal and due to its smallness the (pseudo)scalar correction
can be neglected \cite{C-M,EGPR}.
The invariant amplitudes
$A_{i}^{\mu}{ ^{\; \lambda; \lambda_{\mu}}}({\rm p}) \,$ \cite{ZZS}
in the decay $\pi^{+} \rightarrow \mu^{+} + \nu_{i,\lambda}$, where $\lambda_{\mu} = -1$ or $+1$ is the muon helicity, are related as follows:
\begin{eqnarray}
\label{amplitudes in pion decay -+1}
|A_{i}^{\mu}{ ^{\; +1; +1}}({\rm p})|^2 = |A_{i}^{\mu}{ ^{\; -1; -1}}({\rm p})|^2 \,
\frac{|\varepsilon_{R}|^{2} | U_{\mu i}^{R}|^{2} }{|\varepsilon_{L}|^{2} | U_{\mu i}^{L} |^{2} }\; ,
\end{eqnarray}
where $U_{\alpha i}^{L}$ and  $U_{\alpha i}^{R}$ are the $L$ and $R$ chiral neutrino mixing matrices, which enter into the $CC$ Lagrangian
in the products with the coupling constants $\varepsilon_{L}$ and $\varepsilon_{R}$,
respectively \cite{ZZS}. $U_{\alpha i}^{L}$ is the
Maki-Nakagawa-Sakata-Pontecorvo neutrino mixing matrix \cite{MNSP-1,MNSP-2}.
For relativistic neutrinos, the dependance of the production process on the neutrino
masses can be neglected \cite{Giunti-Kim}.
Then, in the production (P) process, in the center of mass (CM) frame and in
$ |\nu_{i} \rangle_{\lambda}$ basis,
the elements of the general form of the $3\times3$-dimensional nonzero muon neutrino
reduced mass-helicity density matrix
(obtained from the full density matrix by tracing out the other degrees of freedom) are as
follows \cite{OSZ,ZZS}:
\begin{eqnarray}
\label{production matrix - - and  + +}
\varrho^{{\rm P} \mu \; i; \, i'}_{-1; \, -1} \!
= \! \frac{A_{\varepsilon_{L}^{2}}}{\texttt{N}} \, |\varepsilon_{L}|^{2}
U_{\mu i}^{L \ast} U_{\mu i'}^{L}
\; , \;\;\;\;\;
\varrho^{{\rm P} \mu \; i; \, i'}_{+1; \, +1} \! = \!
\frac{A_{\varepsilon_{R}^{2}}}{\texttt{N}} \,  |\varepsilon_{R}|^{2}
U_{\mu i}^{R \ast} U_{\mu i'}^{R} \; ,
\end{eqnarray}
where $\texttt{N} =
A_{\varepsilon_{R}^{2}} |\varepsilon_{R}|^{2} +
A_{\varepsilon_{L}^{2}} |\varepsilon_{L}|^{2} \,$ is the normalization constant
and $A_{\varepsilon_{R}^{2}} = A_{\varepsilon_{L}^{2}}$.
The functions $A_{\varepsilon_{L}^{2}}$ and $A_{\varepsilon_{R}^{2}}$ are the amplitudes
for the $CC$ vector-axial processes, i.e., V-A and V$+$A, respectively.
They depend on
the energies and momenta of the particles in the production process
of the neutrino.
Thus, the density matrix elements are as follows:
\begin{eqnarray}
\label{production matrix - - + +}
\!\!\!\!\!
\varrho^{{\rm P} \mu \; i; \, i'}_{-1; \, -1} \!
= \! \frac{|\varepsilon_{L}|^{2} U_{\mu i}^{L \ast} U_{\mu i'}^{L}}{ |\varepsilon_{R}|^{2} + |\varepsilon_{L}|^{2}} \; , \;\;\;\;\;
\varrho^{{\rm P} \mu \; i; \, i'}_{+1; \, +1} \! = \! \frac{
|\varepsilon_{R}|^{2} U_{\mu i}^{R \ast} U_{\mu i'}^{R} }{ |\varepsilon_{R}|^{2} + |\varepsilon_{L}|^{2}}  \; .
\end{eqnarray}
They constitute the muon neutrino $6 \times 6$-dimensional block diagonal density matrix $\rho^{{\rm P} \mu} \equiv (\varrho^{{\rm P} \mu}_{\lambda ; \,
\lambda'}) =
{\rm diag}((\varrho^{{\rm P} \mu \; i; \, i'}_{-1; \, -1}), (\varrho^{{\rm P} \mu \; i; \, i'}_{+1; \, +1}))$.
We choose $U_{\alpha i}^{R} = U_{\alpha i}^{L}$,
otherwise there is not only the neutrino helicity mixing but also the mass mixing \cite{OSZ}.
%
%
%
Since the density matrix elements (Eq.(\ref{production matrix - - + +}))
depend on the norms of $\varepsilon_{L}$ and $\varepsilon_{R}$,
and not on their phases,
we assume in the ana\-lysis that
these coupling constants are real.
The NP values of $\varepsilon_{L}$ and $\varepsilon_{R}$ can deviate slightly from the SM values 1 and 0,  respectively.
When the constraint $N_{\nu_{+1}}/N_{\nu_{-1}} < 0.002$ is used,
from Eq.(\ref{amplitudes in pion decay -+1}) the bound on the ratio ${\cal R} = |\varepsilon_{R}/\varepsilon_{L}| <
0.0447 \approx 0.045$ results.
Since the Fermi constant constraint
$\varepsilon_{L}^4 + \varepsilon_{R}^4 = 1$ should be also held, we obtain (due to the 4th power)
$|\varepsilon_{R}| <
0.0447  \approx 0.045$
which
constraints the density matrix $(\varrho^{{\rm P}  \mu \; i; \, i'}_{+1; +1})$ of the initial neutrino.

The muon neutrino $\nu_{\mu}$ produced in the process $\pi^{+}  \rightarrow \mu^{+}  + \nu_{\mu}$ is
the relativistic one  (the neutrino energy
$>$ 100 MeV in the Laboratory ({\bf L}) frame). Thus,
the effect of the helicity Wigner rotation is
negligible \cite{OSZ}
resulting in
$\varrho_{\rm {\bf L}}^{{\rm P} \mu}(\vec{{\rm p}}_{\rm {\bf L}}) = \varrho^{{\rm P} \mu}(\vec{{\rm p}})$
for the density matrix in the {\bf L} frame, where $\vec{{\rm p}}_{\rm {\bf L}}$ and $\vec{{\rm p}}$
are the neutrino momenta in the {\bf L} and CM frames, respectively.
Only the neutrino produced in the {\bf L} frame in the forward direction along the $z$-axis
reaches the detector and this axis is chosen as the quantization one \cite{Giunti-Kim}.
After production, the neutrino $\nu_{\mu}$ propagates
in matter and we assume that this is the non-dissipative \cite{DLS} homogeneous medium.
By virtue of quantum mechanical unitarity of the muon-environment time evolution,
the interactions of the entangled muons with their environment cannot affect, in any experiment, the probability of neutrino oscillation that follows the pion decay \cite{Jones}.
Thus, in the
relativistic case, when the distance $z$ and the propagation time $t$ approach the relation
$z=t$
(see, e.g., \cite{Giunti-Kim} for the so-called light-ray approximation and Appendix),
the evolution rule for the neutrino
density matrix
is as follows:
\begin{eqnarray}
\label{evolution rho from T and P}
\!\!
\rho^{\mu}(t=0) \rightarrow \rho^{\mu}(t)  = e^{-i \, {\cal{H}} \,t  }  \rho^{{\rm P} \mu}(t=0) \; e^{i \, {\cal{H}} \,t  } \, ,
\end{eqnarray}
where $\rho^{{\rm P} \mu}$ is the initial density matrix
(\ref{production matrix - - + +}) and ${\cal{H}}$ is the effective Hamiltonian.
%
Although the coherence properties of the neutrino beam resulting from pion decay are
influenced \cite{OSZ} by the initial pion state,
for standard neutrinos no coherence loss
is expected on terrestrial scales \cite{Jones}.
Under the above assumptions, the oscillation probability from $\mu$ to $\beta$ flavour at the detection (D) point
at $z={\rm L}$
is equal to $P_{\mu \rightarrow \beta}({\rm L}) = {\rm Tr}\left[\rho^{\mu}({\rm L}) \hat{P}^{\, \beta}\right]$.
Here
$\hat{P}^{\, \beta} \equiv (\hat{P}^{\, \beta}_{i,i'}) = {\rm diag}((U_{\beta i}^{L}  U^{L \ast}_{\beta i'}), \, (U_{\beta i}^{R}  U^{R \ast}_{\beta i'}))$ is a $6 \times 6$-dimensional block diagonal projection operator
to the $\beta$ flavour direction in the neutrino flavour space \cite{Sz-Z}.
%
%

With three massive and two helicity neutrino states,
${\cal{H}}$ has the $6\times6$-dimensional representation (see, e.g.,
\cite{AZS,ZZS,Dziekuje-za-neutrino-faza}):
\begin{eqnarray}
\label{Eff Ham}
{\cal{H}} = \mathcal{M} + {\cal{H}}_{int} \; ,
\end{eqnarray}
where the $6\times6$-dimensional diagonal matrix $\mathcal{M}$
is the mass term
\cite{AZS,Giunti-Kim}.
Here,
${\cal{H}}_{int}$ is the $6\times6$ matrix representation of the interaction Hamiltonian \cite{ZZS,Dziekuje-za-neutrino-faza} for the coherent neutrino scattering inside the non-dissipative
homogeneous medium \cite{AZS,DLS,Dziekuje-za-neutrino-faza,Dziekuje-za-magnetyzacje}.
%
We will see that, under
the above conditions, data obtained in all earth's oscillation experiments in which the muon neutrinos are produced in the process $\pi^{+}  \rightarrow \mu^{+}  + \nu_{\mu}$
fall into one category of results that together enable the discrimination of the SM from the NP model
(expressed by Eq.(\ref{production matrix - - + +}).)

Note. Usually, the precise knowledge
of the evolution of the neutrino density matrix during oscillation experiments  \cite{AZS,OSZ,ZZS,Kim_Pevsner,Bekman-2002,AZS-conf}
(Appendix),
ruled
by the particular form of the Hamiltonian ${\cal{H}}$ is necessary.
It is the case,
for example, in the consistency analysis \cite{Dziekuje-za-neutrino-faza} of
the values of
parameters of $U_{\alpha i}^{L}$
with the predictions of
(the type of) the Aharonov-Anandan neutrino geometric phase considerations  \cite{sjuk2}.
%
%


\vspace{-3mm}

\section{The SM and NP model discrimination}
\label{The SM and NP model discrimination}


The discrimination of the SM from the NP model presented below takes into account both the problem of its sensitivity and the stability of the $\varepsilon_ {R}$ estimation.
The value of the departure of the purity of the quantum state ${\rm Tr}\left[(\varrho^{\mu})^2\right]$ from 1 \cite{Bengtsson_Zyczkowski},
is a second order effect in the NP parameter $\varepsilon_{R}$ \cite{OSZ}.
In order to
find the distance in the statistical space of distributions, it is convenient to represent the density operator in the spectral-decomposition form, i.e.:
\begin{eqnarray}
\label{spect}
\varrho^{\mu}(z) = \sum_{j=1}^{\aleph} \, p^{j}(z) \, |w_{j}^{\mu}(z)\rangle \langle w_{j}^{\mu}(z)| \, ,
\end{eqnarray}
where $p^{j}(z) \geq 0$ and $|w_{j}^{\mu}(z)\rangle$
are the eigenvalues and (norma\-li\-zed) eigenvectors of
$\varrho^{\mu}(z)$, respectively,
and $\sum_{j=1}^{\aleph} \, p^{j}(z) = 1$,
while the maximal rank of $\varrho^{\mu}$ is equal to $\aleph$.
The NP and SM neutrino quantum states are
given by the density matrices $\varrho^{\mu}_{NP}$
and $\varrho^{\mu}_{SM}$, respectively.
The spectral decomposition of the density matrix
is unique in the sense that for it the von Neumann entropy
$S(\varrho^{\mu}) := - {\rm Tr}(\varrho^{\mu} \ln \varrho^{\mu})$
is equal to the Shannon entropy $S(p)
= - \sum_{j=1}^{\aleph} \, p^{j} \ln p^{j}$ of the probability distribution
$p \equiv \{p^{j}\}_{j=1}^{\aleph}$ \cite{Bengtsson_Zyczkowski}.
The advantage of the spectral-decomposition form
is that via
Fisher-Rao metric (which is related to Shannon entropy \cite{Bengtsson_Zyczkowski}),
the eigenvalues $p^{j}$ enter into the calculation of
the classical lower bound for the variance of the unbiased estimator of a parameter.


The NP effects change the neutrino state
in the course of the oscillation in a different way than
the SM
\cite{Dziekuje-za-neutrino-faza}.
Yet, due to the unitarity of
the evolution
given by Eq.(\ref{evolution rho from T and P}),
the
eigenvalues of the density matrices $\varrho^{\mu}_{SM}$ and $\varrho^{\mu}_{NP}$, which we denote as  $p^{j}_{SM}$ and $p^{j}_{NP}$, respectively, do not vary with $z$.
This happens only if the neutrino evolution remains unitary,
as it is, e.g., in the case of the (constant density) slab approximation
\cite{Giunti-Kim}.
Thus
\begin{eqnarray}
\label{p in 0 and L}
\!\!\!\!\!\!\!\! \!\!\!\!\!\!\!\!
& & p^{j}_{SM} = p^{j}_{SM}(z=0) = p^{j}_{SM}(z = {\rm L})   \; ,
\nonumber \\
\!\!\!\!\!\!\!\! \!\!\!\!\!\!\!\!
& & p^{j}_{NP} = p^{j}_{NP}(z=0) = p^{j}_{NP}(z = {\rm L})  \; , \;\;\; j=1,2,...,\aleph=6 \; .
\end{eqnarray}
It means that,
from the moment of the neutrino production at $z=0$ up to the point of its detection at $z={\rm L}$,
the NP v.s.~SM discrimination reflected in the probability distributions $p_{NP} \! \equiv \! \{p^{j}_{NP}\}$ and $p_{SM} \! \equiv \! \{p^{j}_{SM}\}$ does not change
during
its
propagation.
Therefore,
$p^{j}_{NP}$ and $p^{j}_{SM}$ are the invariants of the neutrino oscillation phenomenon.
This is the reason why
for neutrinos $\nu_{\mu}$ produced in
$\pi^{+}  \rightarrow \mu^{+}  + \nu_{\mu}$
all $\nu_{\mu} \rightarrow \nu_{\mu}$ survival experiments
are integral with each other and
form one general experiment from which all
data can be taken simultaneously.
The limits on the unitary evolution
can appear when, e.g., the sterile neutrinos \cite{Rasmussen}, heavy neutrinos
\cite{Bekman-2002},
decoherence and dissipation  \cite{DLS},
or other phenomena \cite{Rasmussen}, are included, violating the result given by Eq.(\ref{p in 0 and L}).
%
%
In the SM, the produced muon neutrino is in the pure state with helicity $\lambda=-1$, i.e., only one eigenvalue of $\varrho^{\mu}_{SM}$ is nonzero, say $p^{1}_{SM} =1$, and $p^{j}_{SM} =0$ for $j=2,...,6$. In the NP case, the muon neutrino is in
the mixture of two helicity states $\lambda=-1$ and $+1$, i.e., two eigenvalues of $\varrho^{\mu}_{NP}$ are nonzero, say $p^{1}_{NP}$ and $p^{4}_{NP}$ and the others are $p^{j}_{NP}=0$, $j=2,3,5,6$.
From the spectral-decomposition of  $\rho^{{\rm P} \mu}$, Eq.(\ref{production matrix - - + +}),
we obtain numerically (with the accuracy of the expansion coefficients up to the forth decimal place) the truncated series expansion
in $\varepsilon_{R}$ parameter:
\begin{eqnarray}
\label{p on epsRc at L0}
& & p^{4}_{NP} \approx |\varepsilon_{R}|^2 - |\varepsilon_{R}|^4 + 1.4911 \, |\varepsilon_{R}|^6
\;\;\;\;\; \;\;\;\;\; \;\;\;\;\; \;\;\; \;\;\;\;\;\;\;\;\;\;\;\;
{\rm for} \;\;\; {\cal R} < 0.045
\nonumber
\\
& &
p^{1}_{NP} = 1 - p^{4}_{NP}
\;\;\;\;\;\;\; {\rm and} \;\;\;\;\;\;
p^{j}_{NP}=0 \, , \;\;\;\; j=2,3,5,6 \;
\end{eqnarray}
and
$p_{NP} \rightarrow p_{SM}$ for $\varepsilon_{R} \rightarrow 0$.
Through the work, the
symbol of approximate equality
will appear
as a consequence of the approximation occurring in Eq.(\ref{p on epsRc at L0}).



The sensitivity problem is defined as follows:
We are unaware whether
we are sampling from the SM or NP model
distribution.
The sensitivity problem for discrimination of two probability models
is connected with the erroneous identification of the
probability distribution in the $N$-dimensional sampling.
This identification results in
a type II error
of selection of the SM (${\rm H_{0}}$ hypothesis of the statistical
test,) although it is the NP model (${\rm H_{1}}$ hypothesis) which is true.
The
achievable infimum $P_{E}$
of the probability $\beta$ of this error, for
distributions generated by two density matrices, here  $\varrho^{\mu}_{NP}$
and $\varrho^{\mu}_{SM}$, was found by Hiai and Petz \cite{H-P,O-N}:
\begin{eqnarray}
\label{value of PE}
P_{E}(\varrho^{\mu}_{SM},\varrho^{\mu}_{NP}) = e^{- N \,
S(\varrho^{\mu}_{SM}\|\varrho^{\mu}_{NP})} \, ,
\end{eqnarray}
where the
number $N$ of
quantum copies of the system is very large (in principle infinite) and
\begin{eqnarray}
\label{Umegaki}
S(\varrho^{\mu}_{SM}\|\varrho^{\mu}_{NP}) :=
{\rm Tr}[\varrho^{\mu}_{SM}(\ln\varrho^{\mu}_{SM} - \ln\varrho^{\mu}_{NP})] \;
\end{eqnarray}
is the always nonnegative Umegaki quantum relative entropy \cite{Umegaki-1,Umegaki-2}.
$S(\varrho^{\mu}_{SM}\|\varrho^{\mu}_{NP})$ is the measure of how far from each other are
the NP and
SM neutrino quantum states.
By the monotonicity of the relative entropy
it was proven that \cite{Lindblad}
\begin{eqnarray}
\label{relative entropy S1}
\!\!\! \!\!\!\!
S(p_{SM}\|p_{NP})
\leq
S_{1}(\varrho^{\mu}_{SM}\|\varrho^{\mu}_{NP})
\leq
S(\varrho^{\mu}_{SM}\|\varrho^{\mu}_{NP}) \; .
\end{eqnarray}
Here
\vspace{-4mm}
\begin{eqnarray}
\label{Kullback-Leibler}
S(p_{SM}\|p_{NP}) :=
\sum_{j=1}^{6} p^{j}_{SM} \ln\frac{p^{j}_{SM}}{p^{j}_{NP}}
\end{eqnarray}
is the classical Kullback-Leibler relative entropy and
$S_{1}(\varrho^{\mu}_{SM}\|\varrho^{\mu}_{NP})$
is the quantum relative entropy that takes the supremum over all possible
Positive Operator Valued Measures
\cite{Bengtsson_Zyczkowski}.
It should be stressed that from Eq.(\ref{p in 0 and L}) it follows
that the relative entropy $S(p_{SM}\|p_{NP})$
does not vary with $z$, having therefore the same value at the points of the
neutrino production and detection.
By using the relative entropy $S(p_{SM}\|p_{NP})$ in Eq.(\ref{value of PE}) instead of $S(\varrho^{\mu}_{SM}\|\varrho^{\mu}_{NP})$,
the significance of the difference between the NP and SM states is underestimated.
However it gives
the
operationally easier (classical) bound $P_{E}(p_{SM},p_{NP})$
for the calculation of the error probability \cite{Bengtsson_Zyczkowski}:
\begin{eqnarray}
\label{bound on PE}
\!\!\!\!
P_{E}(\varrho^{\mu}_{SM},\varrho^{\mu}_{NP}) \leq
P_{E}(p_{SM},p_{NP})
= e^{- N \, S(p_{SM}\|p_{NP})} \, .
\end{eqnarray}
The relations
(\ref{value of PE}) and (\ref{bound on PE}) are asymptotically strict.

The dependance of $P_{E}(p_{SM},p_{NP})$ on
$\varepsilon_{R}$
for various sample size $N$ is presented in Figure~1.
Since $P_{E}(p_{SM},p_{NP}) \rightarrow  1$ for $\varepsilon_{R} \rightarrow 0$,
thus the smaller
the $\varepsilon_{R}$, the easier the erroneous identification of
the
two models.
To prevent $P_{E}(p_{SM},p_{NP})$
from
increasing
with the decrease of $\varepsilon_{R}$,
the sample size $N$ has to rise.
\begin{figure}[]
\begin{center}
\includegraphics[angle=0,width=100mm,height=70mm]{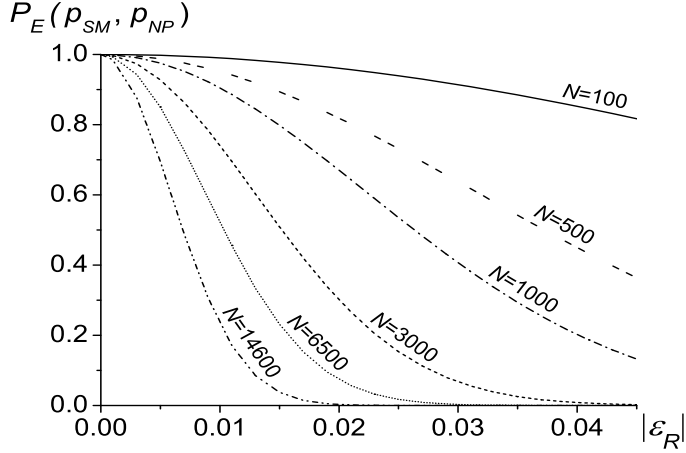}
\end{center}
\vspace{-5mm}
\caption{
The dependance of the upper bound of the erroneous identification
of the model probability distribution $P_{E}(p_{SM},p_{NP})$
on the NP right-chiral $CC$ coupling constant $\varepsilon_{R}$ in the $N$-dimensional sampling.
}
\label{figure1}
\end{figure}
%
%
%

To learn about the stability of the estimation of
$\varepsilon_{R}$,
the lower bound on the variance of its estimator $\hat{\varepsilon}_{R}$
has to be
found.
The relationship between two lower bounds, classical and quantum, will be determined.
It will be shown that the classical lower bound is not smaller than the quantum one, therefore, from the
experimental point of view, the classical bound (which needs the bigger sample) is more restrictive than the quantum bound.
%
%
%

\begin{figure}[]
\begin{center}
\includegraphics[angle=0,width=100mm,height=70mm]{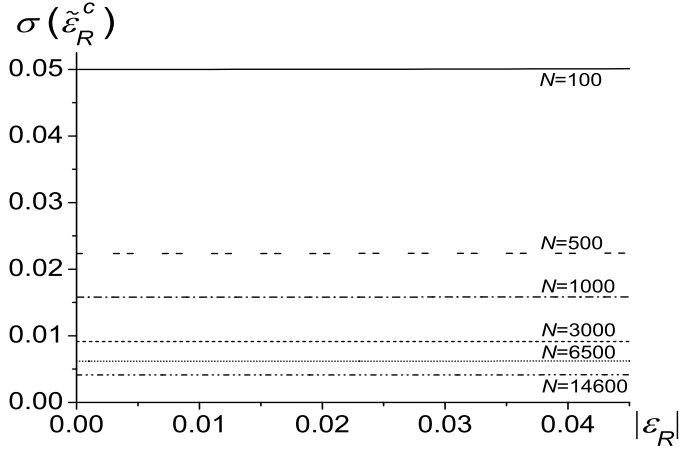}
\end{center}
\vspace{-5mm}
\caption{
The dependance of the lower bound $\sigma(\tilde{\varepsilon}_{R}^{\rm c})$
of the standard error
$\sigma(\hat{\varepsilon}_{R})$ on the NP right-chiral $CC$ coupling constant $\varepsilon_{R}$ in the $N$-dimensional sampling.
For each particular $N$ the quantum lower bound $\sigma(\tilde{\varepsilon}_{R}) = \sqrt{\sigma^{2}(\tilde{\varepsilon}_{R})}$
lies below the corresponding classical lower bound $\sigma(\tilde{\varepsilon}_{R}^{\rm c})$ (see Eq.(\ref{sigma2 inequality})).
}
\label{figure2}
\end{figure}

The classical lower bound is defined as follows: In the classical (c) approach it is the Fisher information on $\varepsilon_{R}$ parameter that has to be  calculated.
In general, a probability distribution $p_{\xi}$ is parameterized by a $n$-dimensional parameter $\xi =  (\xi^{a})_{a=1}^{n}  \in \Xi$, where $\Xi$ is a subset of 
$\mathbb{R}^{n}$.
The Riemannian metric $g_{ab}^{\rm c}$ of the statistical model ${\cal S}=\{ p_{\xi} | \, \xi =  (\xi^{a})_{a=1}^{n} \in \Xi \}$ is called the Fisher-Rao metric \cite{Amari-Nagaoka-book}.
In this paper $\xi$ is reduced to the scalar NP parameter $\varepsilon_{R}$
and the $n=1$-dimensional manifold
${\cal S}=\{  p_{NP}(\varepsilon_{R}) | \, \varepsilon_{R} \in  \langle 0, 1) \}$
is coordinatized by the parameter $\varepsilon_{R}$.
Then, the Fisher-Rao
metric
consists of one component only:
\begin{eqnarray}
\label{Fisher-Rao on subman epsRc - formula}
g_{\varepsilon_{R} \varepsilon_{R}}^{\rm c}
= \sum_{j=1}^{6} p^{j}_{NP} \frac{\partial \ln  p^{j}_{NP}}{\partial \varepsilon_{R}} \frac{\partial \ln p^{j}_{NP}}{\partial \varepsilon_{R}} \; ,
\end{eqnarray}
which, for the distribution $p_{NP}$ given by Eq.(\ref{p on epsRc at L0}), is equal to:
\begin{eqnarray}
\label{Fisher-Rao on subman epsRc}
g_{\varepsilon_{R} \varepsilon_{R}}^{\rm c}(\varepsilon_{R}) \approx
4 \, (1 - 2 \, |\varepsilon_{R}|^2 + 5.4556 \, |\varepsilon_{R}|^4)
\;\;\;\;\;\;\;\; {\rm for} \;\; {\cal R} < 0.045 .
\hspace{-2mm}
\end{eqnarray}
The Fisher information on $\varepsilon_{R}$ in the $N$-dimensional sample is equal to $I_{F}(\varepsilon_{R}) = N \, g_{\varepsilon_{R} \varepsilon_{R}}^{\rm c}$ \cite{Amari-Nagaoka-book}, and
from the scalar
Cram{\'e}r-Rao inequality \cite{Amari-Nagaoka-book}
we obtain, in the classical approach, the lower
bound
$\sigma^{2}(\tilde{\varepsilon}_{R}^{\rm c})$
on the variance of any unbiased estimator $\hat{\varepsilon}_{R}$
of~$\varepsilon_{R}$:
\begin{eqnarray}
\label{sigma2 eps}
\sigma^{2}(\hat{\varepsilon}_{R})
& \geq &
\sigma^{2}(\tilde{\varepsilon}_{R}^{\rm c})
:=
\frac{1}{I_{F}(\varepsilon_{R})} = \frac{1}{N \, g_{\varepsilon_{R} \varepsilon_{R}}^{\rm c}}\nonumber
\\
& \approx &
\frac{1 + 2 \, |\varepsilon_{R}|^2 - 1.4557 \, |\varepsilon_{R}|^4}{4\, N}
\;\;\;\;\;\;\;\;\;\;\;\;\;\;\;\;\;\;\;\;\;
{\rm for} \;\;\; {\cal R} < 0.045 \; .
\end{eqnarray}
Thus, the standard error
$\sigma(\hat{\varepsilon}_{R})
= \sqrt{\sigma^{2}(\hat{\varepsilon}_{R})}
\geq \sigma(\tilde{\varepsilon}_{R}^{\rm c})
= \sqrt{\sigma^{2}(\tilde{\varepsilon}_{R}^{\rm c})}$
$\approx (1 + |\varepsilon_{R}|^2 - 1.2278 \, |\varepsilon_{R}|^4\, )/(2 \sqrt{N})$.
The values of the lower bound $\sigma(\tilde{\varepsilon}_{R}^{\rm c})$
for the standard error $\sigma(\hat{\varepsilon}_{R})$
as the function of $\varepsilon_{R}$ for some $N$ are shown
in Figure~\ref{figure2}.
%
%
%
%
Finally, in the classical approach, the Rao
distance between the distributions $p_{NP}$ and $p_{SM}$
in the statistical model
${\cal S}$
(after applying Eq.(\ref{Fisher-Rao on subman epsRc})), is equal to
\begin{eqnarray}
\label{D-Rao}
D_{Rao}(p_{SM}, p_{NP})
&=&
\int_{0}^{|\varepsilon_{R}|} \sqrt{g_{\varepsilon_{R} \varepsilon_{R}}^{\rm c}(\varepsilon_{R}')} \, d\varepsilon_{R}'
\\
& \approx &
2\, |\varepsilon_{R}| - \frac{2}{3} |\varepsilon_{R}|^3 + 0.8911 \, |\varepsilon_{R}|^5
\;\;\;\;\;\;\;\;  {\rm for} \;\;\; {\cal R} < 0.045 \, .
\nonumber
\end{eqnarray}
%

The bound to the quantum lower bound
is as follows.
Let us consider the distance function
$D(p_{SM}, p_{NP}) := 2 \, \sqrt{S(p_{SM}\| p_{NP})}$, which
for
$\varepsilon_{R} \rightarrow 0$
is consistent with
the Rao distance \cite{Bengtsson_Zyczkowski}.
From Eqs.(\ref{p on epsRc at L0}), (\ref{Kullback-Leibler})
it follows that
\begin{eqnarray}
\label{DpSM-pNP}
D(p_{SM}, p_{NP})
&\approx&
2 |\varepsilon_{R}| - \frac{1}{2} \, |\varepsilon_{R}|^3 + 0.762 \, |\varepsilon_{R}|^5
\;\;\;\;\;\;\;\; {\rm for} \;\;\; {\cal R} < 0.045
\nonumber
\\
& &
\geq D_{Rao}(p_{SM}, p_{NP}) \, .
\end{eqnarray}
The quantum density-operator (DO) distance between the NP and SM neutrino states based on $S(\varrho^{\mu}_{SM}\| \varrho^{\mu}_{NP})$ is given by  $D_{DO}(\varrho^{\mu}_{SM},\varrho^{\mu}_{NP})
:= 2 \, \sqrt{S(\varrho^{\mu}_{SM}\| \varrho^{\mu}_{NP})}$.
Applying the above results
and
Eq.(\ref{relative entropy S1}), we obtain
\begin{eqnarray}
\label{sequence of inequalities}
D_{DO}(\varrho^{\mu}_{SM},\varrho^{\mu}_{NP}) \geq
D(p_{SM}, p_{NP})
\geq D_{Rao}(p_{SM}, p_{NP}) \, .
\end{eqnarray}
Thus, the quantum DO metric
$g^{DO}_{\varepsilon_{R} \varepsilon_{R}}$
(in the square of the line element
$d D_{DO}^{2}(\varrho^{\mu}_{SM},$ $\varrho^{\mu}_{NP})
= g^{DO}_{\varepsilon_{R} \varepsilon_{R}} d\varepsilon_{R}^{2}$),
fulfils the inequality $g^{DO}_{\varepsilon_{R} \varepsilon_{R}}(\varrho^{\mu}_{SM},\varrho^{\mu}_{NP})
\geq g_{\varepsilon_{R} \varepsilon_{R}}^{\rm c}$
and therefore \cite{Braunstein-Caves,Majtey-Lamberti-Prato}:
\begin{eqnarray}
\label{sigma2 inequality}
\sigma^{2}(\tilde{\varepsilon}_{R}^{\rm c})= \frac{1}{N \, g_{\varepsilon_{R} \varepsilon_{R}}^{\rm c}} \geq \sigma^{2}(\tilde{\varepsilon}_{R}):= \frac{1}{N\, g^{DO}_{\varepsilon_{R} \varepsilon_{R}}} \; .
\end{eqnarray}
From Eq.(\ref{sigma2 inequality}) we see that as the classical lower bound $\sigma^{2}(\tilde{\varepsilon}_{R}^{\rm c})$ on
$\sigma^{2}(\hat{\varepsilon}_{R})$ is bigger than the quantum lower bound $\sigma^{2}(\tilde{\varepsilon}_{R})$, hence the quantum estimation is more effective.
Yet, since $\sigma^{2}(\tilde{\varepsilon}_{R}^{\rm c})$ is calculated from Eq.(\ref{Fisher-Rao on subman epsRc - formula})
with the eigenvalues $p^{j}_{NP}$, Eq.(\ref{p on epsRc at L0}), thus,
unlike $\sigma^{2}(\tilde{\varepsilon}_{R})$,
the classical bound depends neither on the relativistic neutrino energy nor on the baseline of the experiment.
Finally, let us note that, from a practical point of view, it appears that if the classical lower bound is experimentally satisfactory (see text below)
then the quantum one, though not designated, is even more powerful.

\vspace{-3mm}

\section{Conclusions}
\label{conclusions}

\vspace{-2mm}

Two model characteristics were evaluated in this paper:
(i)~the model selection one for the sensitivity of the NP-SM
discrimination
with the change of the NP right-chiral $CC$ coupling constant $\varepsilon_{R}$
based
on the upper bound $P_{E}(p_{SM},p_{NP})$ of the erroneous identification of the model probability distribution
$P_{E}(\varrho_{SM},\varrho_{NP})$
and
(ii) the one of the stability of the NP model estimator $\hat{\varepsilon}_{R}$
with the change of $\varepsilon_{R}$ based on
classical Fisher-Rao metric.
The
decay $\pi^{+} \rightarrow \mu^{+} + \nu_{\mu}$ of the high energy pion
followed by the relativistic neutrino unitary propagation
constituted the background for the considerations.

Due to the unitarity of the evolution of the density matrix $\varrho_{NP}$,
its eigenvalues $p^{j}_{NP}$ (Eq.(\ref{p on epsRc at L0})) depend neither on the relativistic neutrino energy nor on the baseline of the experiment.
From this point of view, all $\nu_{\mu} \rightarrow \nu_{\mu}$ survival experiments form one general class.
With these $p^{j}_{NP}(\varepsilon_{R})$, the classical lower bound $\sigma^{2}(\tilde{\varepsilon}_{R}^{c})$  (Eq.(\ref{sigma2 eps})) on the variance of $\hat{\varepsilon}_{R}$ (which is bigger than the quantum lower bound  $\sigma^{2}(\tilde{\varepsilon}_{R})$,
Eq.(\ref{sigma2 inequality})) was calculated.
Thus, with Eq.(\ref{sigma2 eps}) the analysis of the robustness of the
estimation of
$\varepsilon_{R}$ (see text below Eq.(\ref{sigma2 inequality}))
can be
performed
globally, i.e.,
for all production-oscillation (PO)
experiments taken jointly.
To summarize, $N$ in Eq.(\ref{sigma2 eps}) can be taken as the total size of all samples obtained in all survival experiments.

There exists the upper bound (not of the oscillation experiments origin) on the ratio ${\cal R} \equiv |\varepsilon_{R}/\varepsilon_{L}| < 0.045$.
It follows from the analysis of
$\pi^{+} \rightarrow \mu^{+} + \nu_{\mu}$ decay experiments, in which the polarization of the emitted muon was measured \cite{PDG_epsR_2,PDG_epsR_3}.
Eq.(\ref{sigma2 eps}) for
the PO experiments
shows that
for
$|\varepsilon_{R}| \approx 0.045$
to diminish the standard error
$\sigma(\hat{\varepsilon}_{R}) \geq \sigma(\tilde{\varepsilon}_{R}^{\rm c})$
below
$|\varepsilon_{R}|$
value, i.e., for the robust $\varepsilon_{R}$ estimation,
$N \gtrsim 125$ is required.
Then also, for
$N \approx 125$,
$P_{E}(p_{SM},p_{NP}) \approx
0.78$, and the probability of the
erroneous identification
of the NP model with SM
would be high.
Yet, at
the end of 2017
the number $N$ of survival $\nu_{\mu} \rightarrow  \nu_{\mu}$ events in all
$\pi^{+} \rightarrow \mu^{+} + \nu_{\mu}$ experiments
(which form
a combination of T2K, N0vA and mainly MINOS observations)
was already about 6500 \cite{dane-o-survival-1,dane-o-survival-2,dane-o-survival-3,dane-o-survival-4}.
For  $N = 6500$ we obtain $\sigma(\tilde{\varepsilon}_{R}^{\rm c}) \approx 0.006$ and simultaneously the probability $P_{E}(p_{SM},p_{NP}) \approx 2.32
\times 10^{-6}$
is significantly small, leading to good NP-SM discrimination.

In conclusion, if ${\cal R}$ is
only slightly smaller than 0.045, then
both the significant result for the NP-SM discrimination and
robust estimation of the right chiral $CC$ interaction parameter $\varepsilon_{R}$
in neutrino PO experiments have been already reached.
It is anticipated that
2026 will be the first year of the beam operations in the DUNE experiment \cite{DUNE},
which is to result in the observation of more than 7900 $\nu_{\mu}$
survival events
over 3.5 years.
Therefore, in
ten years we will obtain
$N=14600$ survival events,
and even for ${\cal R} = 0.02$
the conventional value
$P_{E}(p_{SM},p_{NP})$ $ <  0.003$ will be
reached, suggesting, if not yet observed, the nonexistence of the right chiral $CC$ neutrino in\-ter\-ac\-tions.
Indeed, on the condition that
the NP model
(hypothesis $H_{1}$) is true and due to Eqs.(\ref{value of PE}) and (\ref{bound on PE}) it follows:
The probability $\beta$ (of erroneous recognition of
the number $N$ of survival $\nu_{\mu} \! \rightarrow  \! \nu_{\mu}$ events
as being predicted by the SM transition rate formula \cite{ZZS} (hypothesis $H_{0}$))
is not bigger than $P_{E}(p_{SM}, p_{NP})$.
Therefore,
even for $N\!=\!14600$ the
selection efficiency $1 - \beta$  for the NP
discovery will be close to $1 -
P_{E}(\varrho^{\mu}_{SM},\varrho^{\mu}_{NP})$ $\geq 1 - 0.003 = 0.997$.
%
This would mean (unless the NP right chiral $CC$ neutrino in\-ter\-ac\-tions
are noticed)
the NP nonexistence,
or at least point to the oddly small value of $\varepsilon_{R}$.

\vspace{-2mm}
\begin{acknowledgements}
This work has been supported by L.J.Ch..\\
%
It has been also supported by
the
Institute of Physics, University of Silesia.
\end{acknowledgements}


\vspace{-3mm}

\appendix
\section*{Appendix: The
density matrix at the detection point}

The effective Hamiltonian ${\cal{H}}$, Eq.(\ref{Eff Ham}),
and neutrino density matrix $\rho^{{\rm P} \mu}$, Eq.(\ref{production matrix - - + +}), have the
$6\times6$-dimensional
matrix re\-presentations.
The diagonalisation of
${\cal{H}}\;$ \cite{AZS,Kim_Pevsner,Bekman-2002} gives
%
${\cal{H}} =  \frac{1}{2 E_{\nu}} W \, {\rm diag} (\tilde{m}_{l}^{2}) \, W^{\dagger}$,
where $W$ is the dia\-gonalising unitary matrix defined by the eigenvectors of $2 E_{\nu} {\cal{H}}$,  and the corresponding real eigenvalues $\tilde{m}_l^2$, $l = 1,...,6 $, are the
neutrino effective squared masses. $E_{\nu}$ is the neutrino energy, neglecting the mass contribution. 
$W \equiv (W_{i \, \lambda; \, l})\equiv (\, {}_{\lambda}\langle \nu_{i}|l \rangle \,)$ defines the transformation from the helicity-mass basis
$|\nu_{i}\rangle_{\lambda}$ $\equiv |{\rm p}, \lambda, i\rangle$
to the eigenvector basis $|l \rangle$ of ${\cal{H}}$.
%
%
%
%
For the relativistic neutrino $\nu_{\mu}$ and in
the non-dissipative homogeneous medium,
from Eq.(\ref{evolution rho from T and P}) it follows that
in $|\nu_{i}\rangle_{\lambda}$ basis
the density matrix at the point $z={\rm L}$ of $\nu_{\mu}$  detection is
\cite{ZZS}:
\begin{eqnarray}
\label{evolution matrix in lamda i}
\varrho^{\mu \; n; \, n'}_{\sigma; \, \sigma'} (t=T) =
\sum_{i \, \lambda} \sum_{i' \lambda'}  \sum_{l,l'}
W_{n \sigma; l} \, W_{i \lambda; l}^{*} \;
\varrho^{{\rm P} \mu \; i; \, i'}_{\lambda; \, \lambda'} (t=0) \,
e^{ - i \, \Delta E_{ll'} \, T}
\,
W_{i' \lambda'; l'} \, W_{n' \sigma'; l'}^{*} \; ,
\vspace{-4mm}
\end{eqnarray}
where $T$ is the time between neutrino production and detection, $\Delta E_{ll'} \equiv E_{l} - E_{l'} = \frac{\Delta \tilde{m}_{ll'}^{2}}{2 E_{\nu}}
= \frac{\tilde{m}_{l}^{2} - \tilde{m}_{l'}^{2}}{2 E_{\nu}}$.
The equality $\varrho_{\rm {\bf L}}^{\mu}(\vec{{\rm p}}_{\rm {\bf L}}) = \varrho^{\mu}(\vec{{\rm p}})$ of the density matrices in the
{\bf L} frame and CM frame is assumed \cite{OSZ}.
Because of the $W$ matrix unitarity, the {\bf L} frame
neutrino density matrix at the detection point is normalized, i.e.,
${\rm Tr}[
\varrho^{\mu}(t = T)] = 1$.
%
%
Eq.(\ref{evolution matrix in lamda i})
is valid in the so-called light-ray approximation $T={\rm L}$  \cite{Giunti-Kim}.
The deviation of
$T$ from the relation $T = {\rm L}$ is experimentally significant if
some corrections $\varepsilon_{ll'}$ \cite{Giunti-Kim} to the oscillation phases $\Delta \phi_{ll'} = |\Delta E_{ll'}| \, T$  are also significant.
As $\varepsilon_{ll'}$ are functions of $\Delta \phi_{ll'}$, this would require
$\Delta \phi_{ll'} \gg 1$  \cite{Giunti-Kim}.
However, for the oscillations to be measurable at all, it is necessary
that $\Delta \phi_{ll'} \sim 1$, in which case the corrections
$\varepsilon_{ll'}$ to
$\Delta \phi_{ll'}$
can be neglected \cite{Giunti-Kim}, validating the
light-ray approximation.
%

Finally,
using $\varrho^{\mu}(t=T)$, Eq.(\ref{evolution matrix in lamda i}), one can also calculate,
e.g., the geometric phase of the $\mu$ flavour neutrino state \cite{Dziekuje-za-neutrino-faza} or
the cross section $\sigma_{\mu \rightarrow \beta}$ for the detection of the $\beta$ flavour neutrino in the {\bf L} frame  \cite{ZZS,AZS-conf}.

\vspace{-5mm}




\begin{thebibliography}{111}


\bibitem{Siringo}
Siringo F.:
Symmetry breaking of the symmetric left-right model without a scalar bidoublet.
Eur. Phys. J. C{\bf 32},  555-559 (2004)


\bibitem{Mohapatra-Pati}
Mohapatra R.N., Pati J.C.:
"Natural" left-right symmetry.
Phys. Rev. D {\bf 11},
2558-2561 (1975)

\bibitem{Zuber}
Zuber K.: Neutrino Physics. Taylor \& Francis Group, New York.
(2004)


\bibitem{Bergmann-Grossman-Nardi}
Bergmann~S., Grossman~Y., Nardi~E.:
Neutrino propagation in matter with general interactions.
Phys. Rev. D {\bf 60}, 093008  (1999)

\bibitem{A-B-Sz-Wudka-Z}
del Aguila~F., de Blas~J., Szafron~R., Wudka~J., Zra{\l}ek~M.:
Evidence for right-handed neutrinos at a neutrino factory.
Phys. Lett. B {\bf 683},
282-288 (2010)


\bibitem{AZS}
del~Aguila~F., Syska~J., Zra{\l}ek~M.:
Impact of right-handed interactions on the propagation of Dirac and Majorana neutrinos
in matter.
Phys. Rev. D {\bf 76},  013007 (2007)


\bibitem{Giunti-Kim}
Giunti~C., Kim~C.W.:
Fundamentals of neutrino physics and astrophysics. Oxford University Press, Oxford (2007)

\bibitem{OSZ}
Ochman~M., Szafron~R., Zra{\l}ek~M.:
Neutrino production state in oscillation phenomena - are they pure or mixed?,
J. Phys. G {\bf 35}, 065003  (2008)


\bibitem{Kuno-Okada}
Kuno~Y.,
Okada~Y.:
Muon decay and physics beyond the standard model.
Rev. Mod. Phys. {\bf 73}, 151-202  (2001)


\bibitem{ZZS}
Syska~J., Zaj{\c a}c~S., Zra{\l}ek~M.,
Neutrino oscillations in the case of general interaction.
Acta Phys. Pol. B {\bf 38}, no.11,  3365-3371 (2007)




\bibitem{PDG_epsR_2}
Fetscher~W.:
Helicity of the $\nu_{\mu}$ in $\pi^{+}$ decay: A comment on the measurement of $P_{\mu}\xi\delta\varrho$ in muon decay.
Phys. Lett.  B  {\bf 140},
117-118 (1984)

%




\bibitem{PDG_epsR_3}
Tanabashi~M. et al., (Particle Data Group):
$\pi^{+}$ - {\it POLARIZATION OF EMITTED} $\mu^{+}$.
Phys. Rev. D {\bf 98}, 030001  (2018).
\\
http://pdglive.lbl.gov/DataBlock.action?node=S008POL











\bibitem{Berman-Kinoshita-1}
Berman~S.M.:
Phys. Rev. Lett. {\bf 1}, 468-469  (1958)

\bibitem{Berman-Kinoshita-2}
Kinoshita~T.:
Radiative corrections to $\pi - e$ decay.
Phys. Rev. Lett. {\bf 2}, 477-480 (1959).

\bibitem{Berman-Kinoshita-3}
Marciano~W.J.,  Sirlin~A.:
Radiative corrections to $\pi_{l2}$ decays.
Phys. Rev. Lett. {\bf 71}, 3629-3632  (1993)


\bibitem{C-M}
Campbell~B.A., Maybury~D.W.,
Constraints on scalar couplings from $\pi^{\pm} \rightarrow l^{\pm} \nu_{l}$.
Nucl. Phys. B {\bf 709}, 419-439  (2005)


\bibitem{EGPR}
Ecker~G., Gasser~J., Pich~A., de~Rafael~E.:
The role of resonances in chiral perturbation theory.
Nucl. Phys. B {\bf 321}, 311-342 (1989)


\bibitem{MNSP-1}
Maki~Z., Nakagawa~M.,
Sakata~S.:
Remarks on the unified model of elementary particles.
Prog. Theor. Phys. {\bf 28}, 870-880 (1962)

\bibitem{MNSP-2}
Pontecorvo~B.:
Neutrino experiments and the problem of conservation of leptonic charge.
JETP {\bf 26}, 984-988 (1968).


\bibitem{DLS}
Dajka~J., Syska~J.,
{\L}uczka~J.:
Geometric phase of neutrino propagating through dissipative matter.
Phys. Rev. D {\bf 83}, 097302 (2011)



\bibitem{Jones}
Jones~B.J.P.:
Dynamical pion collapse and the coherence of conventional neutrino beams.
Phys. Rev. D {\bf 91}, 053002 (2015)


\bibitem{Sz-Z}
Szafron~R., Zra{\l}ek~M.:
Oscillation of Dirac and Majorana neutrinos from muon decay in the case of a general interaction.
Phys. Lett. B {\bf 718}, 113-116 (2012)

\bibitem{Dziekuje-za-neutrino-faza}
Syska~J., Dajka~J.,
{\L}uczka~J.:
Interference phenomenon and geometric phase for Dirac neutrino in $\pi^{+}$ decay.
Phys. Rev. D {\bf 87}, 117302 (2013)


\bibitem{Dziekuje-za-magnetyzacje}
Syska~J.,
Neutrino oscillations in the presence of the crust magnetization.
Nucl. Instr. Methods Phys. Res., Sect. A {\bf 630}, 242-245 (2011)


\bibitem{Kim_Pevsner}
Kim~C.W.,
Pevsner~A.:
Neutrinos in physics and astrophysics.
Contemp. Concepts Phys. Vol. {\bf 8}. Harwood Academic Publishers,
(1993)

\bibitem{Bekman-2002}
Bekman~B., Gluza~J., Holeczek~J., Syska~J., Zra{\l}ek~M.:
Matter effects and CP violating neutrino oscillations with  non-decoupling heavy neutrinos.
Phys. Rev. D {\bf 66}, 093004 (2002)


\bibitem{AZS-conf}
del~Aguila~F., Syska~J., Zra{\l}ek~M.:
Neutrino oscillations beyond the Standard Model.
https://arxiv.org/abs/0809.2759v1


\bibitem{sjuk2}
Tong~D.M., Sj{\"o}qvist~E., Kwek~L.C.,
Oh~C.H.:
Kinematic approach to the mixed state geometric phase in nonunitary evolution.
Phys. Rev. Lett. {\bf 93}, 080405 (2004)
%


\bibitem{Bengtsson_Zyczkowski}
Bengtsson~I., {\.Z}yczkowski~K.:
Geometry of Quantum States. Cambridge University Press, Cambridge (2006).


\bibitem{Rasmussen}
Rasmussen~R.W., Lechner~L., Ackermann~M., Kowalski~M., Winter~W.,
Astrophysical neutrinos flavored with beyond the Standard Model physics.
Phys. Rev. D {\bf 96},
083018 (2017)

\bibitem{H-P}
Hiai~F., Petz~D.:
The proper formula for relative entropy and its asymptotics in quantum probability.
Commun. Math. Phys. {\bf 143}, 99-114 (1991)


\bibitem{O-N}
Ogawa~T., Nagaoka~H.,
Strong converse and Stein's lemma in quantum hypothesis testing.
In Asymptotic Theory of Quantum Statistical Inference,
Selected Papers. Hayashi~M. (ed.),  28-42. Japan Science and Technology Agency \& University of Tokyo, (2005)





%





\bibitem{Umegaki-1}
Umegaki~H.:
Conditional expectation in an operator algebra. IV. Entropy and information.
K{\"o}dai Math. Sem. Rep. {\bf 14}, Nu.2, 59-85 (1962)

\bibitem{Umegaki-2}
Lindblad~G.:
Entropy, information and quantum measurements.
Commun. Math. Phys. {\bf 33},
305-322 (1973)




\bibitem{Lindblad}
Lindblad~G.:
Expectations and entropy inequalities for finite quantum systems.
Commun. Math. Phys. {\bf 39}, 111-119  (1974)


\bibitem{Amari-Nagaoka-book}
Amari~S., Nagaoka~H.:
Methods of information geometry. Translations of mathematical monographs. Vol.{\bf 191}, Oxford University Press, Oxford (2000)


\bibitem{Braunstein-Caves}
Braunstein~S.L.,
Caves~C.M.:
Statistical distance and the geometry of quantum states.
Phys. Rev. Lett. {\bf 72}, 3439-3443 (1994)


\bibitem{Majtey-Lamberti-Prato}
Majtey~A.P., Lamberti~P.W.,
Prato~D.P.:
Jensen-Shannon divergence as a measure of distinguishability between mixed quantum states.
Phys. Rev. A {\bf 72}, 052310 (2005)











\bibitem{dane-o-survival-1}
Adamson~P. et al.,
(MINOS Collaboration):
Measurement of neutrino and antineutrino oscillations using beam
and atmospheric data in MINOS.
Phys. Rev. Lett. {\bf 110}, 251801 (2013)


\bibitem{dane-o-survival-2}
Acero~M.A. et al.,
(NOvA  Collaboration):
New constraints on oscillation parameters from $\nu_{e}$ appearance and $\nu_{\mu}$ disappearance in the NOvA experiment.
Phys. Rev. D {\bf 98}, 032012 (2018)

\bibitem{dane-o-survival-3}
Abe~K. et al.,
(The T2K Collaboration):
Updated T2K measurements of muon neutrino and antineutrino disappearance using
$1.5 \times 10^{21}$ protons on target.
Phys. Rev. D {\bf 96}, 011102(R)  (2017)

\bibitem{dane-o-survival-4}
Carroll~T.J.:
Muon neutrino disappearance measurement at MINOS+.
J. Phys. Conf. Ser.  {\bf 888}, 012161  (2017)


\bibitem{DUNE}
Acciarri~R. at al.,
(The DUNE Collaboration):
Long-Baseline Neutrino Facility (LBNF) and Deep Underground
Neutrino Experiment (DUNE) conceptual design report.
Volume 2: The physics program for DUNE at LBNF.
https://arxiv.org/abs/1512.06148








\end{thebibliography}

\end{document}